\documentclass[letterpaper, 11pt]{article}
\usepackage[nocompress]{cite}
\usepackage{jheppub}

\usepackage{graphicx}
\usepackage{epstopdf}
\usepackage{amsmath, amssymb}

\usepackage{bm}
\usepackage{caption}
\usepackage{subcaption}

\usepackage{multirow} %package for putting different rows in different columns of a table
\usepackage{longtable} %package to allow multipage tables

\usepackage{romannum}%For roman numerals in the title

\usepackage{pdflscape} %For using the enviroment Landscape

\usepackage[dvipsnames]{xcolor}

\usepackage[us,12hr]{datetime} 

\newcommand{\be}{\begin{eqnarray}}
\newcommand{\ee}{\end{eqnarray}}
\newcommand{\nn}{\nonumber}
\newcommand{\bn}{\begin{enumerate}}
\newcommand{\en}{\end{enumerate}}

\def\IZ{\mathbb{Z}}

\def\CA{{\cal A}}
\def\CB{{\cal B}}
\def\CC{{\cal C}}
\def\CD{{\cal D}}

\def\CI{{\cal I}}
\def\CJ{{\cal J}}

\def\CN{{\cal N}}

\def\CQ{{\cal Q}}

\def\CS{{\cal S}}
\def\CT{{\cal T}}

\def\CV{{\cal V}}
\def\CW{{\cal W}}

\def\a{\alpha}

\def\ch{\chi}
\def\half{\frac{1}{2}}

\newcommand{\ket}[1]{\left|{#1}\right\rangle}

\def\Tr{{\rm Tr}}
\def\tr{{\rm Tr}}

\def\PE{\textrm{PE}}

\def\vec#1{\bm{#1}}

\newcommand{\bea}{\begin{eqnarray}}
\newcommand{\eea}{\end{eqnarray}}

\def\IZ{\mathbb{Z}}

\def\CA{{\cal A}} 
\def\CB{{\cal B}}
\def\CC{{\cal C}}
\def\CD{{\cal D}}

\def\CI{{\cal I}}
\def\CJ{{\cal J}}

\def\CN{{\cal N}}

\def\CQ{{\cal Q}}

\def\CS{{\cal S}}
\def\CT{{\cal T}}

\def\CV{{\cal V}}
\def\CW{{\cal W}}

\def\Om{{\mathcal{O}}}
\def\Cm{{\mathcal{C}}}

\def\Bm{{\mathcal{B}}}

\def\Dm{{\mathcal{D}}}
\def\Qm{{\mathcal{Q}}}

\def\a{\alpha}

\def\ch{\chi}
\def\half{\frac{1}{2}}

\def\Tr{{\rm Tr}}
\def\tr{{\rm Tr}}

\def\PE{{\rm PE}}

\allowdisplaybreaks

\def\nn#1{\CN={#1}}

\title{Vanishing OPE Coefficients in 4d $\CN=2$ SCFTs}

\author[a]{Prarit Agarwal,}
\author[b]{Sungjay Lee}
\author[b]{and Jaewon Song}
\affiliation[a]{Department of Physics and Astronomy \& Center for Theoretical Physics\\ Seoul National University, Seoul 151-747, Korea}
\affiliation[b]{School of Physics, Korea Institute for Advanced Study, Seoul 02455, Korea}
\emailAdd{agarwalprarit@gmail.com}
\emailAdd{sjlee@kias.re.kr}
\emailAdd{jsong@kias.re.kr}

\abstract
{
We compute the superconformal characters of various short multiplets in 4d $\CN=2$ superconformal algebra, from which selection rules for operator products are obtained. Combining with the superconformal index, we show that a particular short multiplet appearing in the $n$-fold product of stress-tensor multiplet is absent in the $(A_1, A_{2n})$ Argyres-Douglas (AD) theory. This implies that certain operator product expansion (OPE) coefficients involving this multiplet vanish whenever the central charge $c$ is identical to that of the AD theory. Similarly, by considering the $n$-th power of the current multiplet, we show that a particular short multiplet and OPE coefficients vanish for a class of AD theories with ADE flavor symmetry. 
We also consider the generalized AD theory of type $(A_{k-1}, A_{n-1})$ for coprime $k, n$ and compute its Macdonald index using the associated $W$-algebra under a mild assumption. This allows us to show that a number of short multiplets and OPE coefficients vanish in this theory. We also provide a Mathematica file along with this paper, where we implement the algorithm by Cordova-Dumitrescu-Intriligator to compute the spectrum of $4d$ $\CN=2$ superconformal multiplets as well as their superconformal character.
}

\preprint{SNUTP18-008, KIAS-P18102}

\begin{document}
\maketitle

\section{Introduction}
Any quantum field theory is constrained by its underlying symmetry. The states and operators should transform appropriately under its symmetry algebra. For a superconformal field theory (SCFT), various constraints on the spectrum of operators and the correlation functions can be obtained from the representation theory of the superconformal algebra. 

In this paper, we focus on 4d $\CN=2$ SCFTs and explore the consequences of the kinematic constraints from the superconformal symmetry. The representation theory of superconformal algebra has been analyzed systematically in the literature \cite{DOBREV1985127, Dolan:2002zh, Cordova:2016emh,Cordova:2016xhm}. One of the powerful consequences of superconformal symmetry is that any 4d $\CN=2$ SCFT has a sector (called Schur sector), described by a vertex operator algebra (or chiral algebra) \cite{Beem:2013sza}. This correspondence allows us to compute the correlation functions of `Schur operators' exactly in terms of the associated vertex operator algebra from which universal bounds on the central charges for the global symmetry can be obtained. See also \cite{Lemos:2015orc}. Combined with the selection rules of the operator product expansion (OPE) of two stress-tensor multiplets, the authors of \cite{Liendo:2015ofa} have shown that there is a universal bound for the central charge $ c \ge \frac{11}{30}$ for any interacting $\CN=2$ SCFT. The bound is saturated for the minimal Argyres-Douglas (AD) theory \cite{Argyres:1995jj,Argyres:1995xn} (sometimes called $H_0 = (A_1, A_2)$ theory) when a certain OPE coefficient vanishes. The vanishing of the OPE coefficient comes from the absence of a relevant short multiplet. This has also been verified by analyzing the Macdonald limit of the superconformal index \cite{Gadde:2011uv} of the $(A_1, A_2)$ AD theory \cite{Song:2015wta} upon combining with the selection rule computed in \cite{Liendo:2015ofa}. Our aim in this paper is to generalize the analysis of \cite{Liendo:2015ofa} and \cite{Song:2015wta} to show that various short multiplets disappear (in the Schur sector) for the AD theories so that the corresponding OPE coefficients vanish. 

The superconformal index for the AD theory (and its generalizations \cite{Cecotti:2010fi,Xie:2012hs,Cecotti:2012jx,Cecotti:2013lda, Wang:2015mra}) was computed in \cite{Buican:2015ina,Buican:2015tda,Buican:2017uka,Song:2015wta,Song:2017oew} using the connection between TQFT and the indices of class $\CS$ theories \cite{Gadde:2009kb,Gadde:2011ik,Gadde:2011uv,Gaiotto:2012xa,Rastelli:2014jja}. The associated vertex operator algebra (VOA) for AD theories is rather simple \cite{Cordova:2015nma,Cecotti:2015lab,Xie:2016evu,Buican:2016arp,Song:2017oew,Creutzig:2017qyf,Creutzig:2018lbc}, which allows us to compute the Schur index. It was shown in \cite{Song:2016yfd, Fluder:2017oxm} that it is also possible to obtain the Macdonald index from the VOA, which is more refined than the Schur index. The full superconformal index was obtained via $\CN=1$ gauge theory realizations for (a subset of) AD theories in \cite{Maruyoshi:2016tqk,Maruyoshi:2016aim, Agarwal:2016pjo, Agarwal:2017roi, Benvenuti:2017bpg}. 

Generally, the superconformal index cannot uniquely specify the short multiplet that accounts for a particular term in the index. But as we discuss in this paper, it turns out that AD theory and its generalizations have a rather simple expression for the Schur index (see \cite{Song:2017oew} for example) as a plethystic exponential over a sum of letters in the following form
\begin{align}
 \CI_{\textrm{Schur}}(q) = \PE \left[ \frac{(\textrm{generators}) - (\textrm{relations})}{1-q}\right] \ , 
\end{align} 
so that we are able to identify the generators and relations. Had there been a fermionic generators, it can contribute to the minus sign inside the PE. But in our case, the associated VOA for the AD theories only have bosonic generators so that the minus sign corresponds to the relations. 
%The assumption can be justified by knowing the Macdonald index. 
Once the selection rules for the generators are found, it is possible to uniquely identify the short multiplet that is accounted by the index for the AD theory. 
%
%
%
%\footnote{\color{OliveGreen} Note that, as we also elaborate later in the paper, what we mean by a selection rule here is not identical to the more general operator product expansion in flat space }. 
%
%
%
For example, in the Schur sector (described by the associated VOA and Schur/Macdonald index), we find 
\begin{align}
 \CT^{n+1} \sim \hat{\CC}_{n(\frac{n}{2}, \frac{n}{2})} \sim 0 \quad \textrm{for }(A_1, A_{2n}) \textrm{ theory} , 
\end{align}
where $\CT \equiv \hat{\CC}_{0(0, 0)}$ stands for the stress-tensor multiplet and $\hat{\CC}_{R(j_1, j_2)}$ is a short multiplet in the Schur sector.\footnote{We use the notation of Dolan-Osborn \cite{Dolan:2002zh} throughout this paper.} In general, the right-hand side of the above equation does not vanish. By computing the selection rules, we find that $n$-fold product of stress-tensor multiplet only contains the $\hat{\CC}_{n(\frac{n}{2}, \frac{n}{2})}$ multiplet. Then the (Macdonald) index tells us that the RHS should vanish for the $(A_1, A_{2n})$ theory. 
This relation translates into the existence of a null state at level $2n+2$ for the Virasoro minimal model $M(2, 2n+3)$, which is the associated VOA for this theory. To put it in another way, the existence of a null state in the VOA necessarily implies the existence of a relation in the Schur sector. Therefore certain Schur operator gets lifted (up to recombinations to a long multiplet) from the spectrum of the theory. We use the information from the Macdonald index and the selection rule to pinpoint exactly which operator gets lifted in the Schur sector.

Since any $\CN=2$ SCFT has the stress tensor multiplet that realizes Virasoro algebra, this OPE coefficient vanishes for any theory whenever the central charge $c$ is identical to the value of AD theory. Hence, we have
\begin{align}
 \lambda\left[ \CT, \hat{\CC}_{n-1(\frac{n-1}{2}, \frac{n-1}{2})}, \hat{\CC}_{n(\frac{n}{2}, \frac{n}{2})}\right]^2
  \sim \prod_{i=1}^n (c - c_n) \ , 
\end{align}
with $c_n$ being the central charge of the $(A_1, A_{2n})$ theory $c_n = \frac{n (6 n+5)}{6 (2 n+3)} $ \cite{Aharony:2007dj,Shapere:2008zf}. 

Other than the index, the crucial information we need in our analysis are the selection rules. In order to obtain these, we first compute the superconformal characters for arbitrary long and short superconformal multiplets. This was also studied by \cite{Dobrev:2004tk, Dobrev:2012me}. We implemented the algorithm of Cordova-Dumitrescu-Intriligator \cite{Cordova:2016emh} to construct various supermultiplets from which we compute the characters. Then we take a product of the characters and decompose it in terms of supermultiplets. 
%We do not obtain the full selection rule, but the decomposition as a series expansion turns out to be sufficient for our purpose.
%
%
% 
It is this decomposition of the tensor product of superconformal representations into irreducible ones, that we refer to as our selection rules in this paper. Also, rather than trying to obtain the full decomposition of this product, we find that a series expansion up to a sufficiently high order is enough for our purpose. 

We also study $(A_{k-1}, A_{n-1})$ theories with coprime $k, n$ where the corresponding VOA is given by the $(k, n+k)$ $W_k$-minimal model \cite{Cordova:2015nma}.  We need to know the Macdonald index in addition to the selection rules to unambiguously specify the vanishing short multiplets. The Macdonald index for $k=2$ is known \cite{Song:2015wta,Maruyoshi:2016tqk,Maruyoshi:2016aim}, but not for the more general AD theories. To achieve this, we use the conjectured prescription to obtain the Macdonald index from the associated $W$-algebra \cite{Song:2016yfd, Fluder:2017oxm} as a refined character of the vacuum module. See also \cite{Beem:2017ooy, Bonetti:2018fqz}. By assuming that the generators of the $W$-algebra come only from the scalar primaries, we are able to compute the Macdonald index using the VOA. We explicitly compute the refined vacuum character for the $W_3$-algebra up to level 9. This allows us to find  vanishing short multiplets in these theories.

The organization of this paper is as follows. We review aspects of superconformal representations and explain the method to obtain the selection rules in section \ref{sec:SCA}. 
In section \ref{sec:TnJn}, we consider an $n$-fold product of the stress-tensor and the conserved-currents in $(A_1, A_{2n})$ and $(G^h[n], F)$ theories respectively. We identify the vanishing short multiplets and OPE coefficients for these theories. 
We generalize the discussion to $(A_{k-1}, A_{n-1})$ AD theories in section \ref{sec:AkAn}. We compute the Macdonald index for $k=3$ using the associated $W_3$-algebra, and make a conjecture for general coprime $k, n$. 
Then we conclude in section \ref{sec:Discussion} with a discussion and possible future directions. 
We provide explicit expressions for the characters for a number of short multiplets in appendix \ref{app:char}, and also a number of null states for the $W_3$-algebra in appendix \ref{app:nullstates}. Interested readers can  verify our computations here by using the Mathematica file we provide along with this paper. 

%%%%%%%%%%%%%%%%%%%%%%%%%%%%%%%%%%%%%%%%%%%%%%%
\section{Superconformal characters and selection rules} \label{sec:SCA}

The operators/states in any given $d$-dimensional conformal field theory can be organized into irreducible representations of the $SO(d,2)$ conformal group. The various states in their respective irreducible representations are labeled by their quantum numbers with respect to the compact subgroup, $SO(d) \times SO(2) \subset SO(d,2)$, with $SO(d)$ being the group of Wick-rotated Lorentz transformations and $SO(2)$ being the scale transformation. The operation of lowering/raising the various states is achieved through the action of translation generators $P_\mu$ and special conformal transformations $K_\mu$. The state with the lowest scaling dimension in a given irreducible representation of the conformal group is called the conformal primary. It is annihilated by all $K_\mu$, while the other states, often referred to as the descendants of the conformal primary, are obtained by successively acting with $P_\mu$'s on the conformal primary. It thus follows that a generic conformal multiplet is infinite dimensional. It is customary to label the entire conformal multiplet by the $SO(d) \times SO(2)$-quantum numbers of its conformal primary. 

In a unitary CFT, all the states should have a positive norm. This then places constraints on the quantum numbers that a conformal primary is allowed to have. These are usually expressible in the form of a lower bound on the scaling dimension of a conformal primary with a given $SO(d)$-spin. When this bound is saturated, the norm of certain states in the conformal multiplet becomes zero. Such states and their descendants should be removed from the multiplet, giving us a shorter conformal multiplet. 

When $d=4$, unitarity places the following bounds on conformal primaries with non-zero spin \cite{Mack:1975je}
\begin{align}
\Delta &\geq (j_1 + j_2) +2, \ j_1 \neq 0, \ j_2 \neq 0, \\
\Delta &\geq  (j_1 + j_2) +1, \ j_1 \cdot j_2 = 0, \ j_1 +j_2 \neq 0 \ ,
\end{align} 
where $(j_1,j_2)$ are the $SO(4)$-spin quantum numbers of the conformal primary. When the above bound is saturated, the level-1 descendant with $SO(4)$-spins $(|j_1 -\half|, |j_2 -\half|)$ acquires a zero-norm and drops out. When the conformal primary is a Lorentz scalar \textit{i.e.} $j_1=j_2=0$, there can be zero-norm states at level-2 which then leads to the constraint given by 
\begin{align}
\Delta \geq 1 , \ j_1 =j_2 =0 \ ,
\end{align}  
with the bound being saturated by free scalar fields. 

In $d$-dimensional superconformal field theories, with $3 \leq d \leq 6$, all the operators can be organized into irreducible representations of the superconformal group $\frak S(d, \CN)$. They are 
\begin{align}\label{scftalgs}
\begin{split}
d=3 & \qquad \frak S(3, \CN) = \frak{osp}(\CN| 4) \; \supset \; \frak{so}(3,2)\times \frak{so}(\CN)_R~,  \\[4pt]
d=4 & \hskip11pt \begin{cases} \frak S(4, \CN) = 
\frak{su}(2,2|\CN) \; \supset \; \frak{so}(4,2)\times \frak{su}(\CN)_R\times \frak{u}(1)_R~,\quad \CN \neq 4~, \\ 
\frak S(4, 4) =  \frak{psu}(2,2|4)\supset \frak{so}(4,2)\times \frak{su}(4)_R~,\quad \CN =4~,
\end{cases} 
\\[4pt]
d=5& \qquad \frak S(5,1) = \frak{f}(4) \; \supset \; \frak{so}(5,2)\times \frak{su}(2)_R~, \quad \CN = 1~,  \\[4pt]
d=6 & \qquad \frak S(6,\CN) = \frak{osp}(6,2|\CN ) \; \supset \; \frak{so}(6,2)\times \frak{sp}(2\CN )_R~. 
\end{split}
\end{align}
Thus, the states are now labeled by their quantum numbers with respect to the $R$-symmetry group along with the quantum numbers with respect to $SO(d) \times SO(2) \subset SO(d,2)$. This gives us a superconformal multiplet which is in fact always a collection of a finite number of (non-supersymmetric) conformal multiplets. The operation of lowering/raising is now achieved through the action of Poincare supercharges $\CQ$ and the superconformal supercharges $\CS$ \cite{DOBREV1985127, Minwalla:1997ka}.   
The operator with the lowest scaling dimension is now called the superconformal primary. It is annihilated by all the superconformal supercharges $\CS$. The superconformal algebra then implies that the superconformal primary is also annihilated by all the $K_\mu$'s. The descendants in the superconformal multiplet are obtained by successive action of the Poincare supercharges $\CQ$. Requiring all the superconformal descendants to have a positive norm gives rise to unitarity bounds which when saturated cause some of the superconformal descendants to drop-out, hence giving a short superconformal multiplet. 

In this paper, we will be interested in 4d superconformal field theories with $\nn{2}$ supersymmetry. 
We will further focus on short-multiplets that contain Schur operators, so-called because they contribute to the Schur (and Macdonald) limits of the 4d $\nn{2}$ superconformal index \cite{Kinney:2005ej,Gadde:2011ik,Gadde:2011uv}. The scaling dimension and $U(1)_r$-charge of Schur operators are necessarily given in terms of their $SU(2)_{j_1} \times SU(2)_{j_2} \times SU(2)_R$ quantum numbers as follows:
\begin{align}
\Delta & = 2R + j_1 + j_2, \\
r&=j_2-j_1
\end{align}
 The corresponding supermultiplets are listed in table \ref{tab:SchurMult}.
\renewcommand{\arraystretch}{1.5}
\begin{table}
	\centering
	\begin{tabular}{|l|l|l|l|l|}
		\hline
		Multiplet  & $(\Delta,j_1,j_2,R,r)$ & $\Om_{\rm Schur}$  & $h$ & $r$    \\ 
		\hline 
		$\hat \Bm_R$  & $(2R,0,0,0,R,0)$ &  $\Psi^{11\dots 1}$   &    $R$ &  $0$ \\ 
		\hline
		$\Dm_{R (0, j_2)}$  & $(2R + j_2 +1,0,j_2,R,j_2 +1)$ &    $ \bar{\Qm}^1_{\dot{+}} \Psi^{11\dots 1}_{\dot  + \dots \dot  + }$ &   $R+ j_2 +1$  & $j_2 + \frac{1}{2}$  \\
		\hline
		$\bar \Dm_{R (j_1, 0 )}$  & $(2R + j_1 +1,j_1,0,R,-j_1-1)$ & $ {\Qm}^1_{ +} \Psi^{11\dots 1}_{+   \dots +}$ &     $R+ j_1+1$  & $-j_1 - \frac{1}{2}$  \\
		\hline
		$\hat \Cm_{R (j_1, j_2) }$ & $(2R + j_1 + j_2 + 2,j_1,j_2,R,j_2-j_1)$ &   ${\Qm}^1_{+} \bar{\Qm}^1_{\dot{+}} \Psi^{11\dots 1}_{+   \dots + \, \dot  + \dots \dot  + }$&   
		$R+ j_1 + j_2 +2$ &  $j_2 - j_1$  \\
		\hline
\end{tabular}
\caption{\label{tab:SchurMult} Short multiplets that contain Schur operators using the notation of \cite{Dolan:2002zh}. The second column gives the associated Dynkin labels. The third column indicates where inside the multiplet the Schur operator sits, with $\Psi$ representing the superconformal primary. The last two columns give the $2d$ holomorphic dimension and $r$-charge in terms of $(j_1,j_2,R)$.} 
\end{table}

Of the multiplets listed in table \ref{tab:SchurMult}, we will be particularly interested in the multiplets $\hat \Cm_{0 (0, 0) }$ and $\hat \Bm_1$. 
This is because $\hat \Cm_{0 (0, 0) }$ is the superconformal multiplet formed by the stress-tensor and the $R$-currents of the SCFT. An interacting SCFT with no other decoupled sector contains a unique copy of $\hat \Cm_{0 (0, 0) }$ in its operator spectrum.  Similarly, $\hat \Bm_1$ is the superconformal multiplet containing the conserved currents of the flavor symmetry acting on a given SCFT. The multiplet $\hat \Cm_{0 (j_1, j_2) }$ contains higher spin conserved currents which are never present in an interacting SCFT. Thus the appearance of the $\hat \Cm_{0 (j_1, j_2) }$ superconformal multiplet in the spectrum will indicate the presence of a free-decoupled sector in the theory.

Let us now define the superconformal character of a supermultiplet $V$ as
%with the superconformal primary having quantum numbers $(\Delta, j_1, j_2, R, r)$ as 
%\footnote{Note that we do not insert $(-1)^F$ in our definition. The superconformal character as we define here is non-zero for long multiplets and hence should not be confused with the superconformal index. }
%\be
%\chi_{V(\Delta, j_1, j_2, R, r)}(q, z_1, z_2, y, s) = \Tr_{V} q^{\Delta} z_1 ^{2j_1} z_2 ^{2j_2} y^{2R} s^{-2r} \ ,
%\ee 
\be
\chi_{V}(q, z_1, z_2, y, s) = \Tr_{V} q^{\Delta} z_1 ^{2j_1} z_2 ^{2j_2} y^{2R} s^{-2r} \ ,
\ee 
where the trace is over all the states in the superconformal multiplet (also see \cite{Dobrev:2004tk, Dobrev:2012me}). Here, $\Delta$ is the scaling dimension of the state being traced over while $(2j_1,2j_2)$ give its weights with respect to the $SO(3,1) \simeq SU(2)_{j_1} \times SU(2)_{j_2}$ Lorentz transformations. Similarly, $2R$ is its weight with respect to $SU(2)_R$  and $r$ is its $U(1)_r$ charge. We give explicit expressions of the characters for some of the short multiplets considered in this paper in Appendix \ref{app:char}.

\paragraph{Selection rules}
The superconformal character as defined above can be used to decompose the product of two or more superconformal multiplets into a direct sum over the various possible supermultiplets.
 In order to do so, we expand the product of the characters as a series in the variable $q$. Each monomial in this expansion represents an operator whose scaling dimension and $SU(2)_{j_1} \times SU(2)_{j_2} \times SU(2)_R \times U(1)_r$ charges can be read-off from the monomial. The coefficient of the monomial represents the multiplicity of such operators. The monomial with the lowest power of $q$ in this expansion necessarily represents a superconformal primary. Thus we are guaranteed to have the corresponding superconformal multiplet. We now subtract the character of this superconformal multiplet from our product to obtain a series which start at some higher power of $q$. This then gives us the next supermultiplet that must also be present in the product. We can now repeat the above steps to obtain the list of superconformal multiplets that appear upon decomposing a given product. 

For example, if one considers the product of two stress-tensor multiplets, then up to $\mathcal{O}(q^6)$, the product can be decomposed as 
\begin{align}
\begin{split}
\chi_{\hat{\CC}_{0(0,0)}} \times \chi_{\hat{\CC}_{0(0,0)}} 
& = \chi _{\CA^4_{0,0(0,0)}}+\chi _{\CC_{\frac{1}{2},\frac{1}{2}(\frac{1}{2},0)}}+\chi _{\bar{\CC}_{\frac{1}{2},-\frac{1}{2}(0,\frac{1}{2})}}+\chi _{\CC_{0,1(1,0)}}+\chi _{\bar{\CC}_{0,-1(0,1)}} \\
&~~+\chi_{\hat{\CC}_{1(\frac{1}{2},\frac{1}{2})}}+2\chi _{\CA^5_{0,0(\frac{1}{2},\frac{1}{2})}}+2\chi _{\CC_{\frac{1}{2},\frac{1}{2}(1,\frac{1}{2})}}+2\chi_{\bar{\CC}_{\frac{1}{2},-\frac{1}{2}(\frac{1}{2},1)}} \\
&~~+\chi_{\CC_{0,1(\frac{3}{2},\frac{1}{2})}}+\chi_{\bar{\CC}_{0,-1(\frac{1}{2},\frac{3}{2})}} +\chi _{\hat{\CC}_{1(1,1)}}+3\chi _{\CA^6_{0,0(1,1)}}+ \chi _{\CA^6_{0,0(1,0)}} \\
&~~+\chi_{\CA^6_{0,0(0,1)}}+\ch_{\CA^6_{0,0(0,0)}} + \cdots \ .
\end{split}
\end{align} 
The product of conserved current multiplets decompose into 
\begin{align}
\begin{split}
\chi_{\hat{\CB}_1} \times \chi_{\hat{\CB}_1} &=\chi _{\hat{\CB}_2}+\chi _{\hat{\CC}_{1(0,0)}}+\chi _{\CA^4_{0,0(0,0)}}+\chi _{\hat{\CC}_{1(\frac{1}{2},\frac{1}{2})}}+\chi _{\CA^5_{0,0(\frac{1}{2},\frac{1}{2})}}+\chi _{\hat{\CC}_{1(1,1)}}+\chi _{\CA^6_{0,0(1,1)}}  \\
&~~+\chi _{\CA^6_{0,0(0,0)}}+\chi _{\hat{\CC}_{1(\frac{3}{2},\frac{3}{2})}}+\chi _{\CA^7_{0,0(\frac{3}{2},\frac{3}{2})}}+\chi _{\CA^7_{0,0(\frac{1}{2},\frac{1}{2})}}+\chi _{\hat{\CC}_{1(2,2)}}+\chi _{\CA^8_{0,0(2,2)}}  \\
&~~+\chi _{\CA^8_{0,0(1,1)}}+\chi _{\CA^8_{0,0(0,0)}}+\chi _{\hat{\CC}_{1(\frac{5}{2},\frac{5}{2})}}+\chi _{\CA^9_{0,0(\frac{5}{2},\frac{5}{2})}}+\chi _{\CA^9_{0,0(\frac{3}{2},\frac{3}{2})}}+\chi _{\CA^9_{0,0(\frac{1}{2},\frac{1}{2})}}  \\
&~~+\chi _{\hat{\CC}_{1(3,3)}}+\chi _{\CA^{10}_{0,0(3,3)}}+\chi _{\CA^{10}_{0,0(2,2)}}+\chi _{\CA^{10}_{0,0(1,1)}}+\chi _{\CA^{10}_{0,0(0,0)}} + \cdots \ .
\end{split}
\end{align}
This procedure can be applied to any product of short or long representations to obtain the selection rule. 

Notice that our selection rule is not identical to the more general operator product expansion (such as the ones obtained in \cite{Liendo:2015ofa,Ramirez:2016lyk}) in the flat space. 
We are considering the states on $S^3$ or equivalently the product of operators at the same point. Once we separate the operators in spacetime, we get extra contributions that depend on the distance which eventually reorganize according to the conformal dimension and spin. This is doable in principle, but we do not consider this more general problem. For our purpose, it suffices to focus on the case where all the operators are at the origin since we are interested in the relation between the local operators at the same point. 

\paragraph{Superconformal index for the short multiplets}

The $\mathcal{N}=2$ superconformal index is defined as \cite{Kinney:2005ej}
\begin{align} \label{eq:indexDef}
\begin{split}
\mathcal{I}(p, q, t) &= \Tr (-1)^F p^{j_2 - j_1 -r} q^{j_2 + j_1 -r} t^{R+r} \ .
\end{split}
\end{align}
It receives a non-trivial contribution only from those states that satisfy 
\begin{align}
\label{eq:halfBPS}
\Delta &= 2R +2j_2-r \ .
\end{align} 
These are states that are annihilated by the supercharge, $\widetilde{\CQ}_{1\dot{-}}$. All the other states in an SCFT always appear in Bose-Fermi pairs and hence their net contribution to the index is trivial. The index of long multiplets also evaluates to zero for the same reason. In \cite{Gadde:2011uv}, several simplifying limits of the superconformal index were considered such that non-trivial contributions only come from the states that are annihilated by more than one supercharge. 

In this paper, we will be interested in the limiting cases referred to as the Macdonald and the Schur index. 
The Macdonald index is obtained by taking the limit $p \rightarrow 0$ with $q$ and $t$ fixed in \eqref{eq:indexDef}. Similarly, the Schur index is obtained by taking the limit $t\rightarrow q$ while keeping $p$ arbitrary. Turns out that upon taking the Schur limit, the $p$-dependence drops out and therefore the Schur index is a function of a single fugacity $q$. In both of these limits, the states that contribute non-trivially have to  also satisfy the condition,
\be
\label{eq:quarterBPS}
r = j_2-j_1 \ ,
\ee 
in addition to \eqref{eq:halfBPS}. As mentioned before, these are usually referred to as the Schur operators. The Schur index can also be obtained from the Macdonald index by taking the limit $t\rightarrow q$. 

Let us now compute the Schur and Macdonald indices of the short multiplets $\hat{\CC}_{R(j_1, j_2)}$ and $\hat{\CB}_R$. 
For each of these multiplets, the conformal primary satisfying the Schur conditions is given in the third column of table \ref{tab:SchurMult}. This operator and its descendants arising from the action of $P_{+\dot{+}}$, are the only components of the supermultiplet that contribute to its Macdonald and the Schur indices. For our purposes it will be more useful to redefine the fugacity $t$ in \eqref{eq:indexDef} as $t =q T$. Then the Macdonald index is defined as
\begin{align}
 \CI(q, T) = \tr (-1)^F q^{\Delta-R} T^{R+r} \ , 
\end{align}
where the trace is over the states satisfying \eqref{eq:halfBPS} as well as \eqref{eq:quarterBPS}. 
The Schur index can then be obtained from the Macdonald index by simply setting $T=1$. It then follows that the Macdonald/Schur index for the $\hat{\CC}_{R(j_1, j_2)}$ multiplet is
\begin{align} \label{eq:CRidx}
  \CI_{\hat{\CC}_{R(j_1, j_2)}}(q, T) &= (-1)^{2(j_1+j_2)} \frac{q^{R+2+ j_1 +j_2 }T^{R+1+j_2-j_1}}{1-q} 
   \to 
    (-1)^{2(j_1+j_2)} \frac{q^{R+2+ j_1 +j_2 }}{1-q} \ .
\end{align}
where the arrow refers to taking the Schur limit $T \to 1$. This expression for the index also include other short multiplets in the Schur sector via
\begin{align} \label{eq:SchurIso}
 \hat{\CC}_{R(-\half, j_2)} \simeq \CD_{R+\half(0, j_2)} \ , \quad \hat{\CC}_{R(j_1, -\half)} \simeq \bar{\CD}_{R+\half(j_1, 0)} \ , \quad \hat{\CC}_{R(-\half, -\half)} \simeq \CB_{R+1} \ , 
\end{align}
The index for the $\hat{\CB}_R$ multiplet is given as
\begin{align} \label{eq:BRidx}
 \CI_{\hat{\CB}_R} (q, T) = \frac{q^{R}T^{R}}{1-q} \to  
     \frac{q^{R+1 }}{1-q} \ .
\end{align}

%%%%%%%%%%%%%%%%%%%%%%%%%%%%%%%%%%%%%%%%%%%%%%%%%%%%%%%%
\section{Vanishing short multiplets in $\CT^n$ and $\CJ^n$} \label{sec:TnJn}

\subsection{$\CT^n$ in $(A_1, A_{2n})$ theory}

Let us consider the selection rule for the products of stress tensor multiplets. The stress tensor multiplet is denoted as $\hat\CC_{0(0, 0)}$. It is shown in \cite{Liendo:2015ofa} that the selection rule for the operator-product of two stress tensors is given by 
\begin{align} \label{eq:TT}
\begin{split}
 \hat\CC_{0(0, 0)} \times \hat\CC_{0(0, 0)} \sim \CI + \hat\CC_{0(\frac{\ell}{2}, \frac{\ell}{2})} + \hat\CC_{1(\frac{\ell}{2}, \frac{\ell}{2})} + \ldots  \ . 
\end{split}
\end{align}
The multiplets $\hat\CC_{0(\frac{\ell}{2}, \frac{\ell}{2})}$ are the ones containing the higher spin conserved currents, so they have to be absent in an interacting theory (without any decoupled free sector) \cite{Maldacena:2011jn}. It has been shown in \cite{Liendo:2015ofa} (by using the relation between 4d $\CN=2$ SCFT and chiral algebra \cite{Beem:2013sza}) that the OPE coefficient for the $\hat\CC_{1(\half, \half)}$ can be determined to give
\begin{align}
\left( \lambda [\CT, \CT, \hat\CC_{1(\half, \half)}]\right)^2 = \a \left( 2 - \frac{11}{15 c} \right) \ , 
\end{align}
with $\alpha$ being a positive constant. This yields the lower bound on the central charge $c$ as
\begin{align}
 c \ge \frac{11}{30} \ . 
\end{align}
This value is saturated by the $H_0 = (A_1, A_2)$ Argyres-Douglas theory \cite{Shapere:2008zf}. 

It was also shown in \cite{Song:2015wta} that by combining the Macdonald limit of the superconformal index \cite{Gadde:2011uv} and the selection rule \eqref{eq:TT}, the OPE coefficient $\lambda [\CT, \CT, \hat\CC_{1(\half, \half)} ]$ vanishes. The argument goes as follows. The Macdonald index for the $(A_1, A_2)$ (or $H_0$) AD theory is (conjectured) to be given by 
\begin{align} \label{eq:A1A2idx}
 \CI_{(A_1, A_2)}(q, T) = \PE \left[ \frac{q^2 T - q^4 T^2}{(1-q)(1-q^5 T^2)} + O(q^{11}) \right] \ , 
\end{align}
where $\PE$ stands for the Plethystic exponential. The PE generates the product operators from the `single-trace' type operators. 
This expression has been further verified in \cite{Maruyoshi:2016tqk, Maruyoshi:2016aim,Agarwal:2016pjo, Song:2016yfd, Song:2017oew}. 
The Macdonald index gets contributions only from the short multiplet $\hat{\CC}_{R(j_1, j_2)}$ upon extending $j_{1, 2} \ge -\half$ via relation \eqref{eq:SchurIso} and its index is given in \eqref{eq:CRidx}. 
The stress tensor multiplet $\hat{\CC}_{0(0, 0)}$ and the $\hat{\CC}_{1(\half, \half)}$ multiplet contribute to the index by 
\begin{align}
 \CI_{\hat{\CC}_{0(0, 0)}} = \frac{q^2 T}{1-q} \ , \qquad \CI_{\hat{\CC}_{1(\half, \half)}} = \frac{q^4 T^2 }{1-q} \ . 
\end{align}
From the index of $(A_1, A_2)$ theory \eqref{eq:A1A2idx}, we see that the term contributes to $\hat{\CC}_{1(\half, \half)}$ multiplet is missing. Therefore this multiplet has to be absent and the corresponding OPE coefficient has to vanish. Notice that had we only known the index, we cannot make this statement because the term $\frac{q^4 T^2}{1-q}$ can come from any short multiplet of the form $\hat{\CC}_{\frac{3}{2}-j(\half, j)}$ with $j \in \IZ + \half$. The selection rule \eqref{eq:TT} enables us to unambiguously identify the (lack) of contribution in the index as the one with $j=\half$. 

Essentially, what we are doing is as follows: There is a null state in the associated VOA (which is the Virasoro algebra with $c=-22/5$) given as ($L_{-4} - \frac{3}{5} L_{-2}^2 )\ket{0}$. We use our selection rules to resolve the ambiguities that arise when lifting this VOA null-relation to relations between 4d short-multiplets.  

One might worry that there may be a contribution to the index from a number of short multiplets of the from $\hat{\CC}_{\frac{3}{2}-j(\half, j)}$ so that it contributes to zero (or the minus one inside the PE) in the index. In order for this to happen without vanishing of $\hat{\CC}_{1(\half, \half)}$ multiplet, which has to be there from the OPE selection rule, we need a fermionic generator (in this case, $\hat{\CC}_{\frac{3}{2}(\half, 0)}$ or $\hat{\CC}_{\frac{1}{2}(\half, 1)}$ to account for the $-\frac{q^4 T^2}{1-q}$ term inside the PE of \eqref{eq:A1A2idx}. 
Since the associated VOA for the $(A_1, A_2)$ theory is (conjectured to be) purely bosonic and there is no fermionic generator. Therefore this cannot happen. 

Now, equipped with the character formulae and decomposition in terms of superconformal representations, we work out the selection rule for the higher powers of stress-tensor multiplets. Combining with the index formula for the $(A_1, A_{2n})$ AD theory, let us show that certain OPE coefficient vanishes for this AD theory. The Macdonald index for the $(A_1, A_{2n})$ theory can be written as
\begin{align} \label{eq:A1A2nidx}
 \CI_{(A_1, A_{2n})} (q, T) = \PE \left[ \frac{q^2 T - (q^2 T)^{n+1}}{(1-q)} + O(q^{2n+2}) \right] \ . 
\end{align}
This expression for the index tells us the relation of the form $\CT^{n+1} \sim 0$ if we assume there is no fermionic generator in the associated VOA. We see that the second term inside the numerator comes from a short multiplet of the form $\CC_{\frac{3n}{2}-j(\frac{n}{2}, j)}$. From the character, we find
\begin{align}
 \left( \hat{\CC}_{0(0, 0)} \right)^{n+1} \ni \hat{\CC}_{{n}(\frac{n}{2}, \frac{n}{2})} \ , 
\end{align}
and 
\begin{align}
\left( \hat{\CC}_{0(0, 0)} \right)^{n+1} {\not\owns} ~\hat{\CC}_{{\frac{3n}{2}-j}(\frac{n}{2}, j)} \qquad \mathrm{for }~ j_1 \neq \frac{n}{2} \ .   
\end{align}
Therefore the second term in the index implies the absence of the short multiplet $\hat{\CC}_{{n}(\frac{n}{2}, \frac{n}{2})}$, which proves the conjecture made in \cite{Song:2015wta}. We can further show that the $\hat{\CC}_{n(\frac{n}{2}, \frac{n}{2})}$ multiplet only appears from the OPE of $\hat{\CC}_{{n-1}(\frac{n-1}{2}, \frac{n-1}{2})} \times \CT$. Therefore the following OPE coefficient vanishes:
\begin{align}
 \lambda \left[ \CT, \hat{\CC}_{{n-1}\left(\frac{n-1}{2}, \frac{n-1}{2} \right)}, \hat{\CC}_{{n}\left(\frac{n}{2}, \frac{n}{2}\right)} \right] = 0 \ , \qquad \textrm{for the } (A_1, A_{2n}) \textrm{ theory}. 
\end{align}
More generally, we expect the OPE coefficient above for arbitrary $\CN=2$ SCFT to have a form given by 
\begin{align}
 \lambda \left[ \CT, \hat{\CC}_{{n-1}\left(\frac{n-1}{2}, \frac{n-1}{2} \right)}, \hat{\CC}_{{n}\left(\frac{n}{2}, \frac{n}{2}\right)} \right]^2 \sim \prod_{i=1}^n (c - c_i) \ , 
\end{align}
where $c_n = \frac{n (6 n+5)}{6 (2 n+3)} $ is the central charge of $(A_1, A_{2n})$ theory. This is because the correlation functions of the Schur operators are entirely determined by associated VOA. For the $(A_1, A_{2n})$ theory, VOA is simply given by the Virasoro algebra, which is contained in arbitrary VOA associated to 4d $\CN=2$ SCFT. Hence the OPE coefficients must vanish whenever the central charge is identical to that of the AD theory.  

%%%%%%%%%%%%%%%%%%%%%%%%%%%%%%%%%%%%%%%%%%%%%%
\subsection{$\CJ^n$ in $(G^h [n], F)$ theory}
Let us consider the $n$-fold product of the conserved current multiplet $\hat\CB_1$. By computing the character, we find that 
\begin{align}
 \left(\hat{\CB_1} \right)^n \ni \CB_n \ . 
\end{align}
This can be seen easily without computing the character.  The bottom component of $\hat{\CB}_1$ multiplet has quantum number $(\Delta, j_1, j_2, R, r)=(2, 0, 0, 1, 0)$. The $n$-fold product should contain a state with $(\Delta, j_1, j_2, R, r)=(2n, 0, 0, n, 0)$ satisfying the shortening condition $\Delta = 2R$ for the $\hat{\CB}_R$ multiplet. 

The Argyres-Douglas theory of type $(G^{h}[n], F)$ \cite{Xie:2012hs, Cecotti:2012jx,Cecotti:2013lda,Wang:2015mra} has flavor symmetry $G$ if $(h, n)\equiv 1$. Here $h$ is the dual coxeter number of a Lie algebra $G \in ADE$. The Schur index can be concisely written as \cite{Xie:2016evu, Song:2017oew}
\begin{align}
 \CI_{(G^h [n], F)} (q; \vec{z}) = \PE \left[ \frac{q - q^{h+n}}{(1-q)(1-q^{h+n})} \chi_{\textrm{adj}}(\vec{z})\right] \ , 
\end{align}
where $\chi_{\textrm{adj}}$ refers to the character for the adjoint representation of $G$. The associated chiral algebra is given by a simple affine Kac-Moody algebra $\widehat{\mathfrak{su}}(k)_{k_{2d}}$ with $k_{2d}=-h + \frac{h}{h+n}$. 
The first term in the index comes from the conserved current multiplet $\CJ \equiv \hat\CB_1$. Since any power of $\CJ$ is present in general, the second term with the minus sign means that a certain multiplet contributing to the index that appears in the operator product of $\CJ^{h+n}$ should be absent. 

We find that $\hat\CB_{h+n}$ is the only superconformal multiplet appearing in the $h+n$-fold product of the conserved current, that can account for the second term in the index. Therefore we conclude that $\hat\CB_{n+k}$ in the adjoint sector is absent. In terms of OPE, we find
\begin{align}
 \lambda \left[ \CJ, \hat\CB_{h+n-1}, \hat\CB_{h+n} \right] \Big|_{\textrm{adj}} = 0 \ , 
\end{align}
where $\CB_j$ are in the spin-$j$ representation of $SU(2)_R$. Here we find that only the adjoint sector of this OPE vanishes. For the case of $h+n=2$, we see that the $\hat\CB_2$ short multiplet in the adjoint representation is absent. When $G=SU(N)$, this is precisely the condition the flavor central charge bound $k_{4d} \ge N$ is saturated except for $N=2$ \cite{Beem:2013sza, Lemos:2015orc}.\footnote{For $N=2$, the bound $k_{4d} \ge \frac{8}{3}$ is stronger, and it is saturated for $h+n=3$.} At this value of the flavor central charge, the OPE coefficient $ \lambda \left[ \CJ, \CJ, \hat\CB_{2} \right] \Big|_{\textrm{adj}}$ vanishes.

%%%%%%%%%%%%%%%%%%%%%%%%%%%%%%%%%%%%%%%%%%%%
\section{Vanishing short multiplets in $(A_{k-1}, A_{n-1})$ theory} \label{sec:AkAn}
In this section, we consider generalized Argyres-Douglas theory of type $(A_{k-1}, A_{n-1})$ \cite{Cecotti:2010fi,Xie:2012hs} with $k, n$ being coprime. The Schur index for this theory is known to be given in a concise closed form as \cite{Cordova:2015nma,Song:2015wta,Song:2017oew}
\begin{align} \label{eq:AkAnIdx}
 \CI_{(A_{k-1}, A_{n-1})}(q) = \PE \left[ \frac{\sum_{i=2}^{k} \left(q^i - q^{i+n-1} \right) }{(1-q)(1-q^{k+N})}\right]  \ . 
\end{align} 
This expression is exactly the same as that of the vacuum character of $\CW$-minimal model $\CW(k, k+n)$ \cite{Andrews1999a} with the central charge given as
\begin{align} \label{eq:AkAnc}
 c_{2d} = - \frac{(k-1)(n-1)(k+n+nk)}{k+n} \ , \qquad c_{4d} = -\frac{1}{12} c_{2d} \ . 
\end{align}
This suggests that the associated chiral algebra (or vertex operator algebra) for the $(A_{k-1}, A_{n-1})$ theory is given by the $\CW$-minimal model. Notice that this is only the case when $k, n$ are coprime. See \cite{Buican:2015ina,Buican:2015tda,Buican:2017uka, Creutzig:2017qyf, Creutzig:2018lbc} for the case of $\textrm{gcd}(k, n) \neq 1$. 

This form of the index \eqref{eq:AkAnIdx} suggests that there is a set of generators and relations for the associated vertex operator algebra. The generators of the $W$-algebra are the holomorphic currents of spin $i=2, 3, \ldots, k$. These generators contribute to the term $\frac{q^i}{1-q}$ in the index. In terms of the 4d short multiplets, $i=2$ comes from the $\hat{\CC}_{0(0, 0)}$ multiplet, which contains the stress-energy tensor. 
What about the other terms with $i>2$? Unlike the case of $(A_1, A_{2n})$ theory, neither the full index nor Macdonald index is available. Had we known the Macdonald index, it would be possible to identify the corresponding short multiplet assuming the spins in the bottom component are the same $j_1=j_2$. Since this is not available, we make an educated guess instead. Let us assume that a generator of the chiral algebra comes from the 4d operator without spin. To motivate this assumption, let us consider the case of $k=3$. The short multiplets that can contribute to $\frac{q^3}{1-q}$ to the Schur index are $\hat{\CC}_{R(j_1, j_2)}$ with $R+j_1+j_2=1$ and $j_1 + j_2 \in \IZ_{\ge 0}$. There are only finitely many cases: 
\begin{align}
\begin{split}
&\hat{\CC}_{0(\half, \half)},\quad \hat{\CC}_{0(1, 0)},\quad \hat{\CC}_{0(0, 1)},\quad \hat{\CC}_{1(0, 0)}, \\
&\hat{\CB}_3 \equiv \hat{\CC}_{2(-\half, -\half)},\quad \CD_{\half(0, \frac{3}{2})} \equiv \hat{\CC}_{0(-\half, \frac{3}{2})}, 
\quad \bar{\CD}_{\half(\frac{3}{2}, 0)} \equiv \hat{\CC}_{0(\frac{3}{2}, -\half)}
\end{split}
\end{align}
Among these multiplets $\hat{\CC}_{0(\half, \half)}$ corresponds to the higher-spin conserved current, which should be absent for any interacting theory \cite{Maldacena:2011jn}. Also, $\CD, \bar{\CD}$ type multiplets with spin, called `exotic chiral primaries', are absent for most of the theories we know \cite{Buican:2014qla}. The $\hat{\CB}_R$ contains an operator that parametrizes the Higgs branch, which is not present in the current theory. Therefore if we assume the spins of the generators are the same, which is natural for a Lorentz invariant theory with no preferred direction, we are left with the unique choice $\hat{\CC}_{1(0, 0)}$ for the $q^3$ term in the index. 

We conjecture this assumption to be true in general, which leads us to claim that short multiplets $\hat{\CC}_{i-2(0, 0)}$ are the ones corresponding to the generators. These short multiplets are mapped to the higher-spin currents in the associated $W$-algebra. 
From this assumption, we can compute the Macdonald index from the associated vertex operator algebra for the $(A_{k-1}, A_{n-1})$ theory using the procedure described in \cite{Song:2016yfd, Fluder:2017oxm}. Armed with the Macdonald index and the selection rules, we show that certain OPE coefficients for this theory vanish in a similar way as in section \ref{sec:TnJn}. 

\subsection{Macdonald index of $(A_2, A_{n-1})$ from $W_3$-algebra}

The associated chiral algebra (or VOA) allows us to obtain the Macdonald grading once the $T$ grading for the generators is specified \cite{Song:2016yfd, Fluder:2017oxm}. Let us briefly review how this procedure works. The reader may skip this subsection if one is more interested in the consequences of the Macdonald index. 

The vacuum module of a chiral algebra (or VOA) admits a filtration $ \CV_0 \subset \CV_1 \subset \CV_2 \subset \ldots $ with
\begin{align}
 \CV_k = \textrm{Span} \left\{ X^{(i_1)}_{-n_1} \cdots X^{(i_m)}_{-n_m} \ket{\Omega} : n_1 \ge \cdots \ge n_m , \sum_{j=1}^m w(X^{(i_j)}) \le k \right\} \bigg/ \{ \textrm{null states}\} \ , 
\end{align}
where $\ket{\Omega}$ is the vacuum state and $X^{(i)}$ are the (strong) generators of the chiral algebra with the subscript denoting the mode number in the Laurent expansion. Here the weight $w(X)$ of the generator $X$ is a priori arbitrary from the chiral algebra perspective. We will choose the weights to make connection with the Macdonald index of the 4d theory. 
Once the filtration is given, one can construct an associated vector space as
\begin{align}
 V_{\textrm{gr}} = \bigoplus_{i=0}^{\infty} V_i  \qquad \textrm{with}\quad V_i = \CV_i / \CV_{i-1} \textrm{ and } V_0 = \CV_0\ . 
\end{align}
From here, we define the refined character as 
\begin{align}
 \chi_{\CV}^{\textrm{ref}} (q, T) = \sum_{i \ge 0} \Tr_{V_i} q^{L_0} T^i \ , 
\end{align}
where $L_0$ denotes the Virasoro weight.\footnote{We need to insert $(-1)^F$ to match with the index of the general 4d theory. But for the examples we consider in the current paper, there is no fermionic generator so that we drop it.} 
In short, the refined character introduces additional grading via counting the number of `raising operators' that are needed to reach a particular state with given Virasoro weight. See \cite{Song:2016yfd} for more details. 

We are interested in the refined character for the $W_3$-algebra that is coming from the $(A_2, A_{n-1})$ with $(n, 3) \equiv 1$. Let us compute the refined character for the vacuum module of $W_3$. It has 2 generators $L, W$ with spin 2 and 3, where $L$ is the usual Virasoro generator and $W$ being the higher-spin current. We expect they are coming from the short multiplet $\hat{\CC}_{0(0, 0)}$ and $\hat{\CC}_{1(0, 0)}$ respectively. The Macdonald indices for these multiplets are
\begin{align}
  \CI_{\hat{\CC}_{0(0, 0)}} = \frac{q^2 T}{1-q} \ , \qquad \CI_{\hat{\CC}_{1(0, 0)}} = \frac{q^3 T^2 }{1-q} \ . 
\end{align}
This fixes the weight of the generators as $w(L)=1$ and $w(W)=2$. Therefore the refined character vacuum module when there is no null state besides the vacuum can be written as
\begin{align}
 \chi_{\CV_{\textrm{generic}}}^{\textrm{ref}} (q, T) = \PE \left [ \frac{q^2 T + q^3 T^2}{(1-q)} \right] = \prod_{n \ge 0} \frac{1}{(1-q^{2+n} T)(1-q^{3+n} T^2)} \ . 
\end{align}
We do have null states in the vacuum module corresponding to the $(A_2, A_{n-1})$ Argyres-Douglas theories. 
The correponding Virasoro central charges are $c_{2d}=-\frac{114}{7}, -23, -\frac{186}{5}, -\frac{490}{11}$ for $n=4, 5, 7, 8$ respectively. We list some of the null states explicitly in Appendix \ref{app:nullstates}. 

From this result, we compute the refined character to $q^9$. This translates to the Macdonald index for the corresponding theory. The result is as follows:
\begin{align}
\begin{split}
\CI_{(A_2, A_3)}(q, T) &= 1+q^2 T+q^3 \left(T^2+T\right)+q^4 \left(2 T^2+T\right) \\
 & \quad +q^5 \left(2 T^2+T\right)+q^6 \left(2 T^3+3 T^2+T\right)+q^7 \left(3 T^3+3 T^2+T\right) \\
 & \quad +q^8 \left(T^4+5 T^3+4 T^2+T\right)+q^9 \left(2 T^4+7 T^3+4 T^2+T\right)+O\left(q^{10}\right)
\end{split}
\end{align}
\begin{align}
\begin{split}
\CI_{(A_2, A_4)}(q, T) &= 1+q^2 T+q^3 \left(T^2+T\right)+q^4 \left(2 T^2+T\right)\\
& \quad +q^5 \left(T^3+2 T^2+T\right)+q^6 \left(3 T^3+3 T^2+T\right)+q^7 \left(4 T^3+3 T^2+T\right) \\ 
& \quad +q^8 \left(3 T^4+6 T^3+4 T^2+T\right)+q^9 \left(5 T^4+8 T^3+4 T^2+T\right)+O\left(q^{10}\right)
\end{split}
\end{align}
\begin{align}
\begin{split}
\CI_{(A_2, A_6)}(q, T) &= 1+q^2 T+q^3 \left(T^2+T\right)+q^4 \left(2 T^2+T\right) \\
& \quad +q^5 \left(T^3+2 T^2+T\right) +q^6 \left(T^4+3 T^3+3 T^2+T\right) \\ 
& \quad +q^7 \left(2 T^4+4 T^3+3 T^2+T\right) +q^8 \left(5 T^4+6 T^3+4 T^2+T\right) \\
& \quad +q^9 \left(2 T^5+7 T^4+8 T^3+4 T^2+T\right)+O\left(q^{10}\right)
\end{split}
\end{align}
\begin{align}
\begin{split}
\CI_{(A_2, A_7)}(q, T) &= 1+q^2 T+q^3 \left(T^2+T\right)+q^4 \left(2 T^2+T\right) \\
& \quad +q^5 \left(T^3+2 T^2+T\right)+q^6 \left(T^4+3 T^3+3 T^2+T\right) \\
& \quad +q^7 \left(2 T^4+4 T^3+3 T^2+T\right)+q^8 \left(T^5+5 T^4+6 T^3+4 T^2+T\right) \\
& \quad +q^9 \left(3 T^5+7 T^4+8 T^3+4 T^2+T\right)+O\left(q^{10}\right)
\end{split}
\end{align}
We use this result to find vanishing short multiplets for the $(A_2, A_{n-1})$ theory. 
It would be interesting to come up with a closed-form formula for the Macdonald index (or refined character) for the general $(A_{k-1}, A_{n-1})$ AD theory as was found in \cite{Song:2015wta} for $k=2$.  

\subsection{$(A_2, A_{n-1})$ theory}
\paragraph{$(A_2, A_3)$ theory}
Let us consider the simplest generalized AD theory of type $(A_2, A_3)$. The Schur index reads
\begin{align}
 \CI_{(A_2, A_3)}(q) = \PE \left[ \frac{q^2+q^3 - q^5 - q^6}{(1-q)(1-q^7)} \right] \ , 
\end{align}
so that we have $\hat{\CC}_{0(0, 0)}$ and $\hat{\CC}_{1(0, 0)}$ short multiplets. 
The term $-q^5$ in the numerator inside the PE means that certain short multiplets appear in the OPE of $\hat{\CC}_{0(0, 0)} \times \hat{\CC}_{1(0, 0)}$ should be absent. 
The term $-q^6$ in the Schur index can come from either $\CT^3$ or $\CW^2$ ($\CT \equiv \hat{\CC}_{0(0, 0)}, \CW \equiv \hat{\CC}_{1(0, 0)}$). How can we tell if $-q^6$ comes from $\CT^3$ or $\CW^2$? 
It can be determined from the Macdonald index at $q^6$ order whether this term comes from $\CT^3$ or $\CW^2$. The first would give $-q^6 T^3$ and the latter would give $-q^6 T^4$. 

The Macdonald index can be written as 
\begin{align}
 \CI_{(A_2, A_3)}(q, T) = \PE \left[ \frac{q^2 T + q^3 T^2 - q^5 T^3 - q^6 T^4}{1-q} + O(q^7)\right] \ . 
\end{align}
Therefore we conclude that certain short multiplets appear in the OPE of $\hat{\CC}_{0(0, 0)} \times \hat{\CC}_{1(0, 0)}$ and $\hat{\CC}_{1(0, 0)} \times \hat{\CC}_{1(0, 0)}$ should be absent. In terms of the associated chiral algebra (or VOA), it means that there are null states at level 5 and 6. The term in the index $\frac{q^5 T^3}{1-q}$ comes from the absence of $\hat{\CC}_{R(\half, j_2)}$ with $R+j_2 =\frac{5}{2}$.  
The term $\frac{q^6 T^4}{1-q}$ comes from $R+j_2=\frac{7}{2}$ both with $\half + j_2 \in \IZ_{\ge 0}$. 

The Macdonald index by itself is not enough to fix these charges, but we can fix them by combining with the following selection rules:
\begin{align}
 \hat{\CC}_{0(0, 0)} \times \hat{\CC}_{1(0, 0)} &\ni \hat{\CC}_{2(\half, \half)} \\
 \hat{\CC}_{1(0, 0)} \times \hat{\CC}_{1(0, 0)} &\ni \hat{\CC}_{3(\half, \half)}  
\end{align}
Therefore, we find that $\hat{\CC}_{2(\half, \half)}$ and $\hat{\CC}_{3(\half, \half)}$ multiplets are absent in the $(A_2, A_3)$  AD theory. This also means that the following OPE coefficients vanish for this theory: 
\begin{align}
\lambda\left[ \hat{\CC}_{0(1, 1)} , \hat{\CC}_{1(0, 0)} , \hat{\CC}_{2(\half, \half)}  \right]
= \lambda\left[ \hat{\CC}_{1(0, 0)} , \hat{\CC}_{1(0, 0)} , \hat{\CC}_{3(\half, \half)}  \right] = 0
\end{align}

\paragraph{$(A_2, A_4)$ theory}
Now, let us consider $(A_2, A_4)$ theory. The Schur index is given as
\begin{align}
  \CI_{(A_2, A_4)}(q) = \PE \left[ \frac{q^2+q^3 - q^6 - q^7}{(1-q)(1-q^8)}\right] \ . 
\end{align}
The generators are the same as before: $\hat{\CC}_{0(0, 0)}$ and $\hat{\CC}_{1(0, 0)}$. Now we have the terms $-q^6, -q^7$ coming from the relations among the generators. The term $-q^6$ can come from either $\CT^3$ or $\CW^2$, whereas the term $-q^7$ can only come from $\CT^2 \CW$. The Macdonald index can be written as
\begin{align}
 \CI_{(A_2, A_4)}(q, T) = \PE \left[ \frac{q^2 T + q^3 T^2 - q^6 T^4 - q^7 T^4}{1-q} + O(q^8) \right] \ , 
\end{align}
from which we find that the term $-q^6 T^4$ comes from $\CW^2$. This term suggests that one of the short multiplet of the form $\hat{\CC}_{R(\half, j_2)}$ with $R+j_2 = \frac{7}{2}$ is absent. Likewise, the term $-q^7 T^4$ means that one of $\hat{\CC}_{R(1, j_2)}$ with $R+j_2 = 4$ is absent. The quantum numbers are fixed by the selection rule we compute:
\begin{align}
 \hat{\CC}_{1(0, 0)} \times \hat{\CC}_{1(0, 0)} \quad~~ &\ni \hat{\CC}_{3(\half, \half)} \\
 \hat{\CC}_{0(0, 0)} \times \hat{\CC}_{0(0, 0)} \times \hat{\CC}_{1(0, 0)} &\ni \hat{\CC}_{3(1, 1)}  \label{eq:TTW}
\end{align}
Therefore, $\hat{\CC}_{3(\half, \half)}$ and $\hat{\CC}_{3(1, 1)}$ short multiplets are absent in the $(A_2, A_4)$ theory. 
We can verify that $\hat{\CC}_{3(1, 1)}$ in \eqref{eq:TTW} is coming from $\hat{\CC}_{1(\half, \half)} \times \hat{\CC}_{1(0, 0)}$. Therefore, the following OPE coefficients vanish in $(A_2, A_4)$ theory:
\begin{align}
\lambda\left[ \hat{\CC}_{1(0, 0)} , \hat{\CC}_{1(0, 0)} , \hat{\CC}_{3(\half, \half)}  \right]
= \lambda\left[ \hat{\CC}_{1(\half, \half)} , \hat{\CC}_{1(0, 0)} , \hat{\CC}_{3(1, 1)}  \right] = 0
\end{align}

\paragraph{$(A_2, A_{n-1})$ theory}
We find the Macdonald index for the $(A_2, A_6)$ theory given as
\begin{align}
 \CI_{(A_2, A_6)}(q, T) &= \PE \left[ \frac{q^2 T + q^3 T^2 - q^8 T^5 - q^9 T^6}{1-q} + O(q^{10}) \right] \ . 
\end{align}
This means that we have the relations $\CW^3 \sim 0$ and $\CT \CW^2 \sim 0$ in the Schur sector. 
For the $(A_2, A_7)$ theory, we find
\begin{align}
 \CI_{(A_2, A_7)}(q, T) &= \PE \left[ \frac{q^2 T + q^3 T^2 - q^9 T^6}{1-q} + O(q^{10}) \right] \ . 
\end{align}
We expect there is a term $-q^{10} T^6$ on the numerator inside the PE. If this is the case, we have the relations $\CW^3 \sim 0$ and $\CT^2 \CW^2 \sim 0$ in the Schur sector. 

We notice a pattern here, so that we conjecture the indices for $(A_2, A_{n-1})$ theory with $n=3k+1$ and $n=3k+2$ as
\begin{align} \label{eq:MacA2}
\CI_{(A_2, A_{3k})}(q, T) &= \PE \left[ \frac{q^2 T + q^3 T^2 - q^{3k+2} T^{2k+1} - q^{3k+3} T^{2k+2} }{1-q} + O(q^{3k+4}) \right] \ , \\
\CI_{(A_2, A_{3k+1})}(q, T) &= \PE \left[ \frac{q^2 T + q^3 T^2 - q^{3k+3} T^{2k+2} - q^{3k+4} T^{2k+2} }{1-q} + O(q^{3k+5}) \right] \ . 
\end{align}
This also means that we have the relations
\begin{align}
& \CW^{k+1} \sim 0 , ~~\CT \CW^{k} \sim 0 \quad ~\textrm{for } (A_2, A_{3k}) \textrm{ theory}, \label{eq:A2A3krel} \\
& \CW^{k+1} \sim 0 , ~~\CT^2 \CW^{k} \sim 0 \quad \textrm{for } (A_2, A_{3k+1}) \textrm{ theory}. \label{eq:A2A3kp1rel}
\end{align}
By computing the characters for the short multiplets, we find
\begin{align}
\begin{split}
& ( \hat{\CC}_{1(0, 0)})^{k+1} \ni \hat{\CC}_{2k+1(\frac{k}{2}, \frac{k}{2})} \ ,  \quad 
 \hat{\CC}_{0(0, 0)} \times (\hat{\CC}_{1(0, 0)})^k \ni \hat{\CC}_{2k(\frac{k}{2}, \frac{k}{2})} \ ,  \\
& \qquad  \qquad (\hat{\CC}_{0(0, 0)})^2 \times (\hat{\CC}_{1(0, 0)})^k \ni \hat{\CC}_{2k+1(\frac{k+1}{2}, \frac{k+1}{2})}  \ . 
\end{split}
\end{align}
The short multiplets on the RHS of the selection rule above are the ones contribute appropriately to the $-q^{3k+3}T^{2k+2}$, $-q^{3k+3}T^{2k+2}$ and $-q^{3k+4}T^{2k+2}$ terms in \eqref{eq:MacA2}. 
Therefore, we find $\hat{\CC}_{2k+1(\frac{k}{2}, \frac{k}{2})}$ and $\hat{\CC}_{2k(\frac{k}{2}, \frac{k}{2})}$ multiplets are absent in the $(A_2, A_{3k})$ theory and $\hat{\CC}_{2k+1(\frac{k}{2}, \frac{k}{2})}$, $\hat{\CC}_{2k+1(\frac{k+1}{2}, \frac{k+1}{2})}$ are absent in the $(A_2, A_{3k+1})$ theory. 
As before, we are able find from which channel these short multiplets appear. 
In the end, we find the OPE coefficients of the following form vanish for the $(A_2, A_{3k})$ theory
\begin{align}
\begin{split}
 \lambda\left[ \hat{\CC}_{1(0, 0)} , \hat{\CC}_{2k-1(\frac{k-1}{2}, \frac{k-1}{2})} , \hat{\CC}_{2k+1(\frac{k}{2}, \frac{k}{2})}  \right]
&= \lambda\left[ \hat{\CC}_{0(0, 0)} , \hat{\CC}_{2k-1(\frac{k-1}{2}, \frac{k-1}{2})} , \hat{\CC}_{2k(\frac{k}{2}, \frac{k}{2})}  \right]  \\  
&= \lambda\left[ \hat{\CC}_{1(0, 0)} , \hat{\CC}_{2k-2(\frac{k-1}{2}, \frac{k-1}{2})} , \hat{\CC}_{2k(\frac{k}{2}, \frac{k}{2})}  \right] = 0 \ , 
\end{split}
\end{align}
and the following OPE coefficients vanish for the $(A_2, A_{3k+1})$ theory
\begin{align}
\begin{split}
& \lambda\left[ \hat{\CC}_{1(0, 0)} , \hat{\CC}_{2k-1(\frac{k-1}{2}, \frac{k-1}{2})} , \hat{\CC}_{2k+1(\frac{k}{2}, \frac{k}{2})}  \right] 
= \lambda\left[ \hat{\CC}_{1(\half, \half)} , \hat{\CC}_{2k-1(\frac{k-1}{2}, \frac{k-1}{2})} , \hat{\CC}_{2k+1(\frac{k+1}{2}, \frac{k+1}{2})}  \right] \\ 
& = \lambda\left[ \hat{\CC}_{0(0, 0)} , \hat{\CC}_{2k(\frac{k}{2}, \frac{k}{2})} , \hat{\CC}_{2k+1(\frac{k+1}{2}, \frac{k+1}{2})}  \right]=\lambda\left[ \hat{\CC}_{1(0, 0)} , \hat{\CC}_{2k-1(\frac{k}{2}, \frac{k}{2})} , \hat{\CC}_{2k+1(\frac{k+1}{2}, \frac{k+1}{2})}  \right] = 0 \ .  
\end{split}
\end{align}
In addition, any OPE coefficients involving absent short-multiplets should vanish. 

%%%%%%%%%%%
\subsection{$(A_{k-1}, A_{n-1})$ theory}
Let us write $n=mk+r$ with $0< r < k$. Then the Schur index can be written as
\begin{align} \label{eq:AkAnSchur}
 \CI_{(A_{k-1}, A_{mk+r-1})} (q) = \PE \left[ \frac{q^2 + \cdots q^k - (q^{mk+r+1} + \cdots + q^{mk+k+r-1})}{(1-q)(1-q^{(m+1)k+r})}  \right] \ . 
\end{align}
We do not know the Macdonald for the general $(A_{k-1}, A_{n-1})$ theory. It should be possible to obtain the Macdonald index from the refined vacuum character of the $W_k$-algebra, but we leave it as a future work. Instead, we give a conjectural expression based on previous computations. 

Let us specialize to the case of $n=mk+1$. Then we conjecture the Macdonald index can be written as
\begin{align}
\begin{split}
& \CI_{(A_{k-1}, A_{mk})} (q, T) = \PE \Bigg[\frac{q^2 T + q^3 T^2 + \cdots + q^k T^{k-1} }{1-q} \\
 & \qquad \quad - \frac{ q^2 T (q^k T^{k-1})^m + q^3 T^2 (q^k T^{k-1})^m  +\cdots + (q^k T^{k-1})^{m+1}  }{1-q} + O(q^{(m+1)k-2}) \Bigg] \ .
\end{split}
\end{align}
If we write the short multiplets as $\CW_i \equiv \hat{\CC}_{i-2(0, 0)}$, the index implies we have the relations in the Schur sector as 
\begin{align}
\CW_i (\CW_k)^m \sim 0 \quad \textrm{ for each }  i=2, \ldots, k \ .
\end{align}
This is a natural generalization of the relation \eqref{eq:A2A3krel} for the $(A_2, A_{3m})$ theory. 
By combining the index and superconformal characters, we see that the short multiplets $\hat{\CC}_{mk-m+i-2(\frac{m}{2}, \frac{m}{2})}$ with $i=2, \ldots k$ must vanish. Therefore, any OPE coefficients involving this short multiplet vanishes including
\begin{align}
\lambda\left[ \hat{\CC}_{k-2(0, 0)} , \hat{\CC}_{mk-m+i-2(\frac{m}{2}, \frac{m}{2})} , \hat{\CC}_{mk-m+k-2(\frac{m}{2}, \frac{m}{2})}  \right] = 0 \ , 
\end{align}
for the $(A_{k-1}, A_{km})$ theory. 

Likewise, we can write a conjectural Macdonald index for the case of $r=k-1$ as
\begin{align}
\begin{split}
& \CI_{(A_{k-1}, A_{mk+k-2})} (q, T) = \PE \Bigg[\frac{q^2 T + q^3 T^2 + \cdots + q^k T^{k-1} }{1-q} \\
 & \qquad \qquad  - \frac{ (q^k T^{k-1})^{m} q^{k-1} T^{k-2} \left( q T +  q^{2} T + \cdots + q^{k-1} T^{k-2} \right) }{1-q} + O(q^{(m+1)k-2}) \Bigg] \ .
\end{split}
\end{align}
The index implies the relations 
\begin{align}
(\CW_k)^{m+1} \sim 0 \ , \quad \CW_i \CW_{k-1} (\CW_k)^m \sim 0 \quad \textrm{ for } i=2, \ldots, {k-1} \ . 
\end{align}
which is a natural generalization of the relation \eqref{eq:A2A3kp1rel} for the $(A_2, A_{3m+1})$ theory. 

For a general $n=mk+r$, we conjecture the relations for the generators as
\begin{align}
\begin{array}{cl}
 (\CW_k)^m \CW_{k-1} \CW_i \sim 0 \quad & \textrm{for } i=2, 3, \ldots, r \ , \\
 (\CW_k)^m \CW_j \sim 0  & \textrm{for }j=r+1, r+2, \ldots, k  \ . 
\end{array}
\end{align}
This gives us the Macdonald index as 
\begin{align}
\begin{split}
\CI_{(A_{k-1}, A_{mk+r-1})}(q, T) &= \PE \Bigg[\frac{q^2 T + q^3 T^2 + \cdots + q^k T^{k-1} }{1-q} \\
 & \qquad - \frac{(q^k T^{k-1})^{m} q^{k-1} T^{k-2}\left( q^2 T + \cdots q^r T^{r-1} \right)}{1-q} \\
  & \qquad - \frac{ (q^k T^{k-1})^{m}\left( q^{r+1} T^r + \cdots q^k T^{k-1} \right) }{1-q} + O(q^{(m+1)k-r}) \Bigg] \ , 
\end{split}
\end{align}
which reduces to \eqref{eq:AkAnSchur} upon taking $T \to 1$. It would be interesting to verify this is indeed the correct index. For any theories of type $(A_{k-1}, A_{mk+r-1})$ with coprime $k, n$, the relation $(\CW_k)^m \sim 0$ is satisfied. From this and the selection rule, we find 
\begin{align}
 \lambda \left[ \hat{\CC}_{k-2(0, 0)}, \hat{\CC}_{mk-m+i-2(\frac{m}{2}, \frac{m}{2})} , \hat{\CC}_{mk-m+k-2(\frac{m}{2}, \frac{m}{2})} \right] = 0 \quad \textrm{ for } i=2, \ldots r \ . 
\end{align}
More generally, we expect the above OPE coefficients vanish when the central charge is given as \eqref{eq:AkAnc}. 

%%%%%%%%%%%%%%%%%%%%%%%%%%%%%%%%%%%%%
\section{Conclusion} \label{sec:Discussion}
In this paper, we have shown that certain short multiplets and the OPE coefficients involving products of stress-tensor vanish for a class of Argyres-Douglas type theories. From this, we were able to argue that OPE coefficients for a general 4d $\CN=2$ SCFT vanish when the central charge $c$ is identical to the AD theory. 

We would like to mention a couple of interesting directions to pursue. First, we expect that the vanishing of specific short multiplets and the OPE coefficients can be a crucial input data for choosing the target in the superconformal bootstrap program \cite{Beem:2014zpa, Lemos:2015awa, Lemos:2016xke,Cornagliotto:2017dup, Cornagliotto:2017snu}. For the AD theories, the Schur sector is completely fixed by the Virasoro or $W$-algebra. It would be interesting to see if this data is enough to fix the CFT to a certain degree.\footnote{There exist two different theories that give rise to the same associated VOA. This happens whenever the two CFTs are related by discrete gauging.} 

It should also be possible to find the exact OPE coefficients for the vanishing ones we considered in the current paper. Once the superconformal block with arbitrary spin is known, we can decompose the correlators of Schur operators that we can obtain from the chiral algebra in terms of the blocks. Then the relevant OPE coefficients can be determined as in \cite{Beem:2013sza}. 

Finally, our analysis in section \ref{sec:AkAn} relied on a few conjectures necessary to compute Macdonald index. It would be interesting to prove that the generators of the associated VOA come from the spin-less primaries. Any other independent computation of the Macdonald index will corroborate our prescription.

%%%%%%%%%%%%%%%%%%%%%%%%%%%%%%%%%%%%%%%%%%%%%
\begin{acknowledgments}
J.S. would like to thank UESTC and Kavli IPMU for hospitality where the paper was finalized. The work of P.A. is supported in part by Samsung Science and Technology Foundation under Project Number SSTF-BA1402-08, in part by National Research Foundation of Korea grant number 2018R1A2B6004914 and in part by the Korea Research Fellowship Program through the National Research Foundation of Korea funded by the Ministry of Science and ICT, grant number 2016H1D3A1938054.
The work of S.L. is supported in part by the National Research Foundation of Korea (NRF) Grant NRF-2017R1C1B1011440.
The work of J.S. is supported in part by the National Research Foundation of Korea (NRF) Grant NRF- 2017R1D1A1B06034369.

\end{acknowledgments}

%%%%%%%%%%%%%%%%%%%%%%%%%%%%%%%%%%%%%%%%%%%%%%%%%
\appendix

\section{Characters for a number of short multiplets} \label{app:char}
Let us write down characters for a number of important short multiplets we consider in this paper. 
For the 4d $\nn{2}$ stress-tensor multiplet $\hat{\CC}_{0(0,0)}$, the above superconformal character evaluates to   
\begin{align}
\begin{split}
\chi_{\hat{\CC}_{0(0,0)}} &= q^2+q^{5/2} \left(\frac{\chi _2^{y} \chi _2^{z_1}}{s}+s \chi _2^{y} \chi _2^{z_2}\right) \\ 
&~~+ q^3 \left(2 \chi _2^{z_1}\chi _2^{z_2}+\chi _2^{z_1} \chi _2^{z_2} \chi_3^y+\frac{\chi _3^{z_1}}{s^2}+s^2 \chi _3^{z_2}\right) + \hdots \ ,
\end{split}
\end{align}
where, $\chi_d^{z_1}$ represents character of the $d$-dimensional irreducible representation of $SU(2)_{j_1}$, with analogous interpretation for $\chi_d^{z_2}$ and $\chi_d^{y}$. 
Similarly, the character of the short multiplet labeled as $\hat{\CC}_{1(0, 0)}$ is given by
\begin{align}
\begin{split}
\chi_{\hat{\CC}_{1(0,0)}} &= q^4 \chi _3^y+q^{9/2} \left(\frac{\chi _2^{y}\chi _2^{z_1}}{s}+s \chi _2^{y}\chi _2^{z_2}+\frac{\chi _2^{z_1} \chi _4^y}{s}+s \chi _2^{z_2} \chi _4^y\right) \\ 
&~~ + q^5 \bigg(\frac{1}{s^2}+s^2+\chi _2^{z_1}\chi _2^{z_2}+\frac{\chi _3^y}{s^2}+s^2 \chi _3^y+3 \chi _2^{z_1}\chi _2^{z_2} \chi _3^y+\frac{\chi _3^{y}\chi _3^{z_1}}{s^2} \\
&~~~~~~~~~~+  s^2 \chi _3^{y}\chi _3^{z_2}+\chi _2^{z_1}\chi _2^{z_2} \chi _5^y\bigg) + \hdots \ .
\end{split}
\end{align}
The character of the short multiplet $\hat{\CC}_{1(\half, \half)}$ is given by
\begin{align}
\begin{split}
\chi_{\hat{\CC}_{1(\half,\half)}} &= q^5 \chi _2^{z_1} \chi _2^{z_2}\chi _3^y  \\ 
&~~+ q^{\frac{11}{2}} \left(s \chi _2^{y}\chi _2^{z_1}+\frac{\chi _2^{y}\chi _2^{z_2}}{s}+\frac{\chi _2^{y}\chi _2^{z_2} \chi
	_3^{z_1}}{s}+s \chi _2^{y}\chi _2^{z_1} \chi _3^{z_2}+\frac{\chi _2^{z_2} \chi _3^{z_1} \chi _4^y}{s}+s \chi _2^{z_1} \chi _3^{z_2} \chi
_4^y\right) \\
&~~+ q^6 \bigg(1+\frac{\chi _2^{z_1}\chi _2^{z_2}}{s^2}+s^2\chi _2^{z_1}\chi _2^{z_2}+\chi _3^y+\frac{\chi _2^{z_1}\chi _2^{z_2} \chi_3^y}{s^2}+s^2 \chi _2^{z_1}\chi _2^{z_2} \chi _3^y+\chi _3^{z_1} \\
&~~~~~~~~~~ + 2 \chi _3^{y}\chi _3^{z_1}+\chi _3^{z_2}+2 \chi _3^{y}\chi _3^{z_2}+\chi _3^{z_1}\chi _3^{z_2}+3\chi _3^{y}\chi _3^{z_1}\chi _3^{z_2}+\frac{\chi _2^{z_2} \chi _3^y \chi _4^{z_1}}{s^2}  \\
&~~~~~~~~~~ + s^2 \chi _2^{z_1} \chi _3^y \chi _4^{z_2}+\chi _3^{z_1}\chi _3^{z_2} \chi _5^y\bigg) + \hdots \ .
\end{split}
\end{align}
The character of short multiplet $\hat{\CC}_{2(\half, \half)}$ is given by
\begin{align}
\begin{split}
\chi_{\hat{\CC}_{2(\half,\half)}} &= q^7 \chi _2^{z_1} \chi _2^{z_2} \chi _5^y  \\
&~ + q^{\frac{15}{2}} \left(s \chi _2^{z_1} \chi _4^y+\frac{\chi _2^{z_2} \chi _4^y}{s}+\frac{\chi _2^{z_2} \chi _3^{z_1} \chi _4^y}{s}+s \chi _2^{z_1} \chi _3^{z_2}
\chi _4^y+\frac{\chi _2^{z_2} \chi _3^{z_1} \chi _6^y}{s}+s \chi _2^{z_1} \chi _3^{z_2} \chi _6^y\right) \\
&~+ q^8 \Big(\chi _3^y+\frac{\chi _2^{z_1} \chi _2^{z_2}  \chi _3^y}{s^2}+s^2 \chi _2^{z_1} \chi _2^{z_2}  \chi _3^y+\chi _3^{z_1} \chi _3^{y}+\chi _3^{y} \chi _3^{z_2}+\chi _3^{y}\chi _3^{z_1} \chi _3^{z_2}+\chi _5^y  \\
&~~~~~~~~~ + \frac{\chi _2^{z_1} \chi _2^{z_2}  \chi _5^y}{s^2}+s^2 \chi _2^{z_1} \chi _2^{z_2}  \chi _5^y+2 \chi _3^{z_1} \chi _5^y+2
\chi _3^{z_2} \chi _5^y+3 \chi _3^{z_1} \chi _3^{z_2}  \chi _5^y+\frac{\chi _2^{z_2} \chi _4^{z_1} \chi _5^y}{s^2} \\
&~~~~~~~~~~~~~~ +s^2 \chi _2^{z_1} \chi _4^{z_2} \chi _5^y+\chi _3^{z_1} \chi _3^{z_2} \chi _7^y\Big) + \hdots \ .
\end{split}
\end{align}
The character for the $\hat{\CC}_{2(1, 1)}$ is given by
\begin{align}
\begin{split}
\chi_{\hat{\CC}_{2(1,1)}} &= q^8 \chi _3^{z_1}\chi _3^{z_2} \chi _5^y   \\ 
&~~+ q^{\frac{17}{2}} \left(s \chi _2^{z_2} \chi _3^{z_1} \chi _4^y+\frac{\chi _2^{z_1} \chi _3^{z_2} \chi_4^y}{s}+\frac{\chi _3^{z_2} \chi _4^{y+z_1}}{s}+s \chi_3^{z_1} \chi _4^{y+z_2}+\frac{\chi _3^{z_2} \chi _4^{z_1} \chi_6^y}{s}+s \chi _3^{z_1} \chi _4^{z_2} \chi _6^y\right) \\ 
&~~+ q^9 \bigg(\chi _2^{z_1}\chi _2^{z_2} \chi _3^y+\frac{\chi_3^{y}\chi _3^{z_1}\chi _3^{z_2}}{s^2}+s^2 \chi_3^{y}\chi _3^{z_1}\chi _3^{z_2}+\chi _2^{z_2} \chi _3^y \chi _4^{z_1}+\chi _2^{z_1} \chi _3^y \chi _4^{z_2}+\chi_3^y \chi _4^{z_1}\chi _4^{z_2} \\
&~~~~~~~~~+ \chi _2^{z_1}\chi _2^{z_2} \chi _5^y+\frac{\chi _3^{z_1}\chi _3^{z_2} \chi _5^y}{s^2}+s^2 \chi _3^{z_1}\chi _3^{z_2} \chi _5^y+2\chi _2^{z_2} \chi _4^{z_1} \chi _5^y+2 \chi _2^{z_1} \chi _4^{z_2} \chi _5^y \\
&~~~~~~~~~+ 3 \chi _4^{z_1}\chi _4^{z_2} \chi _5^y+\frac{\chi _3^{z_2}\chi _5^{y}\chi _5^{z_1}}{s^2}+s^2 \chi _3^{z_1} \chi _5^{y}\chi _5^{z_2}+\chi _4^{z_1}\chi _4^{z_2} \chi _7^y\bigg) + \hdots  \ .
\end{split}
\end{align}
The character of the conserved current multiplet $\hat{\CB}_1$ is given by
\begin{align}
\chi_{\hat{\CB}_{1}} = q^2 \chi _3^y+ q^{5/2} \left(\frac{\chi _2^{y}\chi _2^{z_1}}{s}+s \chi _2^{y}\chi _2^{z_2}\right)+ q^3 \left(\frac{1}{s^2}+s^2+2\chi _2^{z_1}\chi _2^{z_2}+\chi _2^{z_1}\chi _2^{z_2} \chi _3^y\right) + \hdots \ , 
\end{align}
and the character of the $\hat{\CB}_2$ multiplet is given by
\begin{align}
\begin{split}
\chi_{\hat{\CB}_{2}} &= q^4 \chi _5^y+q^{9/2} \left(\frac{\chi _2^{z_1} \chi _4^y}{s}+s \chi _2^{z_2} \chi _4^y\right) \\
& \qquad +  q^5 \left(\frac{\chi _3^y}{s^2}+s^2 \chi _3^y+\chi _2^{z_1}\chi _2^{z_2}\chi _3^y+\chi _2^{z_1}\chi _2^{z_2} \chi _5^y\right) +\hdots \ .
\end{split}
\end{align}

%%%%%%%%%%%%%%%%%%%%%%%%%%%%%%%%%%%%%%%%%%%%%%%%%%%%%%
\section{Null states of $W_3$ vacuum module} \label{app:nullstates}
The commutation relations for the $W_3$ algebra is given as 
\begin{align}
\begin{split}
 [L_m, L_n] &= (m-n) L_{m+n} + \frac{c}{12} m(m^2 - 1) \delta_{m+n, 0} \ , \\
 [L_m, W_n] &= (2m-n) W_{m+n} \ , \\
 [W_m, W_n] &= \frac{c}{360}m(m^2-1)(m^2-4) \delta_{m+n, 0} + \frac{16}{22+5c} (m-n) \Lambda_{m+n} \\ 
  %& \quad + (m-n)\left( \frac{(m+n+2)(m+n+3)}{15} - \frac{(m+2)(n+2)}{6} \right) L_{m+n} \ . 
  & \qquad \qquad + \frac{1}{30} (m-n) (2m^2+2n^2 - mn - 8) L_{m+n} \ . 
\end{split}
\end{align}
Let us define the vacuum state $\ket{\Omega, c}$ as
\begin{align}
 L_n \ket{\Omega, c} = W_n \ket{\Omega, c} = 0 \quad n \ge 0 \ . 
\end{align}
The vacuum module can be constructed by acting negative modes of $L, W$ on the vacuum state. One can easily see that the states $L_{-1} \ket{\Omega, c}$, $W_{-1}\ket{\Omega, c}$ and $W_{-2}\ket{\Omega, c}$ have zero norm. 
We list some of the non-trivial null states in the vacuum module of the $W_3$ algebra that we use to compute the refined character. 
\paragraph{Level 5} 
\begin{align}
\left( W_{-3}  L_{-2} - \frac{10}{7} W_{-5} \right) \left| \Omega, c=-\frac{114}{7}\right\rangle
\end{align}

\paragraph{Level 6}
\begin{align}
& \left( \frac{102}{49} L_{-6} -\frac{24}{7} L_{-4} L_{-2} -\frac{6}{7} (L_{-3})^2 +(L_{-2})^3 - \frac{39}{7} (W_{-3})^2 \right) \left| \Omega, c=-\frac{114}{7} \right\rangle \\
& \left( \frac{27}{4} L_{-6} - \frac{21}{4} L_{-4} L_{-2} - \frac{51}{32} (L_{-3})^2  + (L_{-2})^3 - 
 \frac{279}{16} (W_{-3})^2 \right) \left| \Omega, c=-23 \right\rangle
\end{align}

\paragraph{Level 7}
\begin{align}
& \left( - \frac{5}{8} W_{-7} -\frac{3}{8} L_{-4} W_{-3} - \frac{11}{4} W_{-5} L_{-2} - \frac{3}{16} W_{-4} L_{-3} + 
 W_{-3}  L_{-2}  L_{-2}  \right) \ket{\Omega, c=-23} \\
& \left( L_{-3} L_{-2}^2-\frac{9}{4} L_{-5} L_{-2}+\frac{27 L_{-7}}{8}-\frac{15}{8} L_{-4} L_{-3}-\frac{93}{8} W_{-4} W_{-3} \right) \ket{\Omega, c=-23}
\end{align}

\paragraph{Level 8}
\begin{align}
\begin{split}
&\bigg( \frac{1413}{5} L_{-8} - 15 L_{-6} L_{-2} - \frac{531}{20} L_{-5} L_{-3} - 
 \frac{1011}{100} (L_{-4})^2 + \frac{1599}{10} W_{-5} W_{-3} + \frac{369}{40} (W_{-4})^2  \\ 
&\qquad  - \frac{48}{5} L_{-4} (L_{-2})^2 - 
 \frac{15}{4} (L_{-3})^2 L_{-2}  +  (L_{-2})^4 - \frac{123}{2} (W_{-3})^2  L_{-2} \bigg) \ket{\Omega, c=-\frac{186}{5}}
 \end{split}
 \end{align}

\paragraph{Level 9}
\begin{align}
\begin{split}
& \bigg( -\frac{123}{8} L_{-3} W_{-3}^2-\frac{123}{4} L_{-2} W_{-4} W_{-3}-\frac{1}{16} 15 L_{-3}^3+L_{-2}^3 L_{-3}\\ 
& \quad -\frac{141}{5} L_{-6} L_{-3}  -\frac{111}{20} L_{-4} L_{-2} L_{-3}-\frac{27}{10} L_{-5} L_{-2}^2+\frac{11061 L_{-9}}{25} \\
& \quad -\frac{1587}{50} L_{-5} L_{-4} -\frac{309}{10} L_{-7} L_{-2} +\frac{246}{5} W_{-5} W_{-4}+\frac{4797}{40} W_{-6} W_{-3} \bigg) \ket{\Omega, c=-\frac{186}{5}} 
\end{split} \\
\begin{split}
& \bigg( L_{-2}^3 W_{-3}-\frac{42}{11} L_{-2}^2 W_{-5}-\frac{877}{121} L_{-2} W_{-7}-\frac{9}{11} L_{-3} L_{-2} W_{-4} \\
& \quad -\frac{51}{11} L_{-4} L_{-2} W_{-3} +\frac{777}{121} L_{-3} W_{-6}+\frac{596}{121} L_{-4} W_{-5}+\frac{38}{121} L_{-5} W_{-4} \\ 
& \qquad -\frac{15}{11} L_{-3}^2 W_{-3}  +\frac{948}{121} L_{-6} W_{-3}+\frac{27620}{1331}  W_{-9} -\frac{138}{11} W_{-3}^3 \bigg) \ket{\Omega, c=-\frac{490}{11}}
\end{split}
\end{align}

\bibliographystyle{jhep}
\bibliography{refs}

\end{document}